\documentclass[10pt,conference]{IEEEtran}
\usepackage{amsmath,amssymb}
\usepackage{graphicx}
\usepackage{mathrsfs}
\usepackage{tikz}
\usepackage{booktabs}
\usepackage{floatflt}
\usepackage{enumerate}
\usepackage{psfrag}	
\usepackage{array}
\usepackage{multirow,hhline}
\usepackage{exscale}
\usepackage{color}
\usepackage{colortbl}
\usepackage{epsfig,subfigure}
\usepackage{cite}
\usepackage{amsthm}
\newcommand{\mat}[1]{\ensuremath{\mathbf{#1}}}
\renewcommand{\vec}[1]{\ensuremath{\mathbf{#1}}}
\newtheorem{theorem}{Theorem}
\newtheorem{lemma}{Lemma}

\newtheorem{definition}{Definition}

\begin{document}

% \IEEEoverridecommandlockouts

% \title{The DOF of the $K$ user Interference Channel with a Cognitive Relay}
% \author{Anas Chaaban and Aydin Sezgin\\
% Emmy-Noether Research Group on Wireless Networks\footnote{
% This work is supported by the German Research Foundation, Deutsche
% Forschungsgemeinschaft (DFG), Germany, under grant SE 1697/3.}\\
% Institute of Telecommunications and Applied Information Theory\\
% Ulm University, 89081, Ulm, Germany\\
% Email: {anas.chaaban@uni-ulm.de, aydin.sezgin@uni-ulm.de}}

\IEEEoverridecommandlockouts
\title{The Capacity Region of the Linear Shift Deterministic Y-Channel}
\author{
\IEEEauthorblockN{Anas Chaaban and Aydin Sezgin}
% \IEEEauthorblockA{}
\thanks{%
Emmy-Noether Research Group on Wireless Networks, Ulm University, 89081 Ulm, Germany. Email: {anas.chaaban@uni-ulm.de, aydin.sezgin@uni-ulm.de}.
The work of A. Chaaban and A. Sezgin is supported by the German Research Foundation, Deutsche
Forschungsgemeinschaft (DFG), Germany, under grant SE 1697/3.
}
}

\maketitle

% \footnotetext[1]{a}

\begin{abstract}
The linear shift deterministic Y-channel is studied. That is, we have three users and one relay, where each user wishes to broadcast one message to each other user via the relay, resulting in a multi-way relaying setup. The cut-set bounds for this setup are shown to be not sufficient to characterize its capacity region. New upper bounds are derived, which when combined with the cut-set bounds provide an outer bound on the capacity region. It is shown that this outer bound is achievable, and as a result, the capacity region of the linear shift deterministic Y-channel is characterized.
\end{abstract}

\section{Introduction}
Multi-way communications is a situation where nodes communicate with each other in a bi-directional manner. The first multi-way communications studied setup is the two-way channel studied by Shannon \cite{Shannon_TWC} where 2 nodes communicate with each other, and each has a message to deliver to the other node. Several extensions of this setup were considered. One such extension is obtained by combining relaying and multi-way communications to obtain the so-called multi-way relay channel. For instance, in the two-way relay channel (or the bi-directional relay channel), two nodes communicate with each other via a relay. This setup was introduced in \cite{RankovWittneben}, and further studied for instance in \cite{KimDevroyeMitranTarokh,GunduzTuncelNayak}.

An approximation for the capacity of wireless networks can be obtained by using the deterministic model introduced in \cite{AvestimehrDiggaviTse}. Interestingly, by obtaining the capacity of the deterministic model of some wireless network, we can draw conclusions on the capacity region of its Gaussian model. For instance, the capacity region of the deterministic 2-way relay channel was obtained in \cite{AvestimehrSezginTse} and used to obtain the capacity region of the Gaussian 2-way relay channel within a constant gap. In \cite{AvestimehrKhajehnejadSezginHassibi}, the deterministic multi-pair bi-directional relay network was studied and its capacity was obtained, and in \cite{SezginKhajehnejadAvestimehrHassibi}, the approximate capacity of the Gaussian counterpart was obtained.

The multi-way relay channel was studied in \cite{GunduzYenerGoldsmithPoor} where in this case, users communicate in a multi-way manner by multicasting a message to other users via the relay. A broadcast variant of this multi-way relaying setup, the so called Y-channel, was considered in \cite{LeeLim} in its multiple-input multiple-output variant. That is, 3 MIMO nodes communicate via a MIMO relay, and each node has two messages to broadcast to the other nodes.
 
In this paper, we consider a linear shift deterministic Y-channel. We assume that all nodes are full-duplex, and that the channels are reciprocal. We provide upper bounds on the achievable rates, and show that the provided bounds are tighter than the cut-set bounds. This is contrary to the deterministic two-way relay channel \cite{AvestimehrSezginTse} and the multi-pair bi-directional relay channel \cite{AvestimehrKhajehnejadSezginHassibi} where the cut-set bounds characterize the capacity region. This is similar to the multicast bi-directional communications scenario considered in \cite{MokhtarMohassebNafieElGamal} where the cut-set bounds are not tight. We then show that the outer bound, provided by the collection of obtained upper bounds, is achievable. The capacity achieving scheme combines bi-directional communication, uni-directional communication, and a scheme that we call "cyclic communication". Consequently, we characterize the capacity region of the linear shift deterministic Y-channel.

The rest on the paper is organized as follows. The system model is described in section \ref{Model}. Upper bounds are given in section \ref{UpperBounds} and the capacity achieving transmit strategy is described in section \ref{Achievability}. We summarize the results in section \ref{Summary}. We use the following notation throughout the paper. Scalars are represented by normal font, vectors and matrices by bold face font, and sets by calligraphic font. For instance, $x$, $\vec{x}$, and $\mathcal{X}$ are a scalar, a vector, and a set respectively. $\vec{x}^n$ denotes a sequence of $n$-vectors $(\vec{x}_1,\dots,\vec{x}_n)$.

\section{System Model}
\label{Model}
\begin{figure}[t]
\centering
\includegraphics[width=0.65\columnwidth]{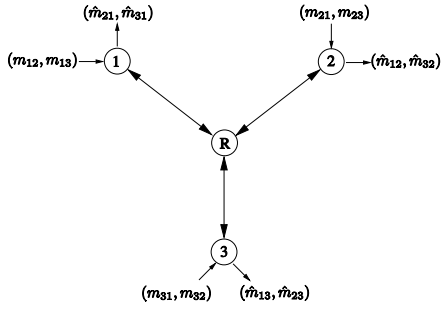}
\caption{The Y-channel.}
\label{Fig:Model}
\end{figure}

The Y-channel, shown in Fig. \ref{Fig:Model}, is a multi-way relaying setup where 3 users communicate with each other in a bi-directional manner via a relay. Since direct links between the users are missing, the relay is essential for communication. Each user of the Y-channel wants to broadcast two messages, one to each other user, via the relay. All nodes are assumed to be full duplex and to have the same power constraint.

Message $m_{jk}$ from user $j$ to user $k$, $j,k\in\{1,2,3\}$ is uniformly distributed over the set $\mathcal{M}_{jk}\triangleq\{1,\dots,2^{nR_{jk}}\}$ for all $j\neq k$ where $R_{jk}\in\mathbb{R}_+$. User $j$ encodes his messages into a codeword $\vec{x}_j^n$ and transmits this codeword. The $i$th symbol of $\vec{x}_j^n$ is $x_{ji}=f_{ji}(m_{jk},m_{jl},\vec{y}_j^{i-1})$, and $\vec{y}_j^{i-1}$ are all received symbols at user $j$ until time instant $i-1$. The relay listens to the transmission of the users, constructs $\vec{x}_r^n$ where $x_{ri}=f_{ri}(\vec{y}_r^{i-1})$ from its received signal $y_r$, and sends it back to the users. User $j$ tries to decode $(\hat{m}_{kj},\hat{m}_{lj})=g_j(\vec{y}_j^n,m_{jk},m_{jl})$ where $g_j(.)$ is the decoding function, and an error occurs if $(\hat{m}_{kj},\hat{m}_{lj})\neq({m}_{kj},{m}_{lj})$. The collection of message sets, encoders, and decoders defines a code for the Y-channel.

\begin{definition}
A rate tuple $(R_{12},R_{13},R_{21},R_{23},R_{31},R_{32})$ denoted $\vec{R}$ is said to be achievable if there exist a sequence of codes such that the average error probability can be made arbitrarily small by increasing $n$. The set of all achievable rate tuples is the capacity region denoted $\mathcal{C}$.
\end{definition}

\subsection{The linear shift deterministic Y-channel}
\begin{figure}[t]
\centering
\includegraphics[width=0.9\columnwidth]{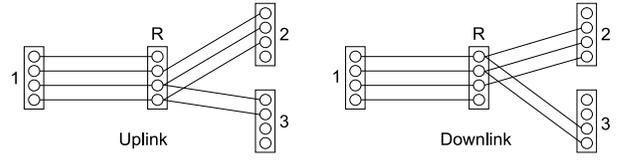}
\caption{A DYC$(4,3,2)$. In the uplink, the relay receives all symbols from user 1 that are above the noise level. Users 2 and 3 have weaker channels, and thus symbols at low levels arrive below the noise level and are clipped. Similarly in the downlink, the symbols at the lower levels at the relay are clipped at receivers 2 and 3.}
\label{Fig:DYC_Model}
\end{figure}

The channel gains of a Gaussian Y-channel are modeled by non-negative integer gains $n_j$, $j\in\{1,2,3\}$, in the linear shift deterministic Y-channel (DYC) (see \cite{AvestimehrDiggaviTse}). We assume that the channel is reciprocal, i.e. the channel gain from one user to the relay is the same as that from the relay to this user. Without loss of generality, we can assume that 
\begin{align}
\label{Ordering}
n_1\geq n_2\geq n_3.
\end{align}
The transmit signal of user $j$ is a $q$-dimensional binary vector $\vec{x}_{ji}\in\mathbb{F}_2^q$ where $q=\max_j\{n_j\}$. The received signal at each node, $\vec{y}_{ri},\vec{y}_{ji}\in\mathbb{F}_2^q$, is a deterministic function of the transmit signals, modeled by a shift of the transmit signal. That is
\begin{align}
\label{IO1}
\vec{y}_{ri}=\sum_{j=1}^3 \mat{S}^{q-n_j}\vec{x}_{ji},\quad\text{and}\quad\vec{y}_{ji}=\mat{S}^{q-n_j}\vec{x}_{ri}
\end{align}
where $\vec{S}^{q-n_j}$ is a $q\times q$ shift matrix that, when multiplied with a vector, shifts its rows downwards by $n_j$ positions. All operations are done in $\mathbb{F}_2$. Notice how $n_j$ models the gain of the channel, and how the effect of noise is modeled as clipping symbols of the transmit signals in lower positions. A linear shift deterministic Y-channel with levels $n_1$, $n_2$ and $n_3$ is denoted DYC$(n_1,n_2,n_3)$. A DYC$(4,3,2)$ is shown in Fig. \ref{Fig:DYC_Model}. A line between two circles in Fig. \ref{Fig:DYC_Model} represents a bit-pipe between these two levels, which models (\ref{IO1}).

\section{Upper bounds}
\label{UpperBounds}

In this section, we provide some upper bounds on the achievable rates of the DYC. We start with the single rate bounds given by $R_{jk}\leq\min\{n_j,n_k\}$. We also have the cut-set bounds as follows
\begin{align}
\label{CSB1}
R_{jk}+R_{jl}&\leq \min\{n_j,\max\{n_k,n_l\}\}\\
\label{CSB2}
R_{kj}+R_{lj}&\leq \min\{n_j,\max\{n_k,n_l\}\}.
\end{align}
for all distinct $j,k,l\in\{1,2,3\}$. These bounds already provide an outer bound on the capacity region $\mathcal{C}$. 

In many deterministic bi-directional communication setups, the cut-set bounds characterize the whole capacity region \cite{AvestimehrKhajehnejadSezginHassibi,AvestimehrSezginTse}. However, in some deterministic bi-directional setups, the cut-set bounds are not enough for characterizing the capacity region \cite{MokhtarMohassebNafieElGamal} and further bounds are required. The DYC belongs to the latter case. In fact, many bounds from the cut-set bounds will be shown to be redundant due to the bounds we provide next. The proofs of the following lemmas are omitted due to space limitation. The proof follows a genie-aided approach similar to that in \cite[Lemmas 1 and 2]{ChaabanSezgin_YC_SC}.

\begin{lemma}
\label{FromRelay}
The achievable rates in the DYC must satisfy
\begin{align}
\label{BFR}
\hspace{-0.2cm}R_{kj}+R_{lj}+R_{kl}&\leq \max\{n_j,n_l\},\ \ \forall \{j,k,l\}=\{1,2,3\}.
\end{align}
\end{lemma}

\begin{lemma}
\label{GUB}
The achievable rates in the DYC must satisfy
\begin{align}
\label{BTR}
\hspace{-0.2cm}R_{kj}+R_{lj}+R_{kl}&\leq \max\{n_k,n_l\},\ \ \forall \{j,k,l\}=\{1,2,3\}.
\end{align}
\end{lemma}

The following theorem is obtained from Lemmas \ref{FromRelay} and \ref{GUB} by considering all $\{j,k,l\}=\{1,2,3\}$.
\begin{theorem}
The achievable rates in the DYC satisfy
\begin{align}
\label{TRB1}
R_{12}+R_{32}+R_{13}&\leq n_2\\
\label{TRB2}
R_{12}+R_{32}+R_{31}&\leq n_1\\
\label{TRB3}
R_{21}+R_{31}+R_{32}&\leq n_2\\
\label{TRB4}
R_{21}+R_{31}+R_{23}&\leq n_2\\
\label{TRB5}
R_{13}+R_{23}+R_{12}&\leq n_2\\
\label{TRB6}
R_{13}+R_{23}+R_{21}&\leq n_1.
\end{align}
\end{theorem}

% Instead of bounding the sum of 2 rates, the bounds \eqref{TRB1}-\eqref{TRB6} provide bounds on the sum of three rates, which make some of the cut-set bounds redundant as we see next. 
By evaluating the single rate bounds and the bounds in (\ref{CSB1}) and (\ref{CSB2}) using (\ref{Ordering}), we notice that the individual rates are redundant given the cut-set bounds:
\begin{align}
\label{CS1}
R_{12}+R_{13}\leq n_2, & \quad  R_{21}+R_{31}\leq n_2,\\
\label{CS2}
R_{21}+R_{23}\leq n_2, & \quad  R_{32}+R_{12}\leq n_2,\\
\label{CS3}
R_{31}+R_{32}\leq n_3, & \quad  R_{13}+R_{23}\leq n_3.
\end{align}
Moreover, the cut-set bounds in (\ref{CS1}) are redundant given (\ref{TRB1}) and (\ref{TRB3}). Similarly, the cut-set bounds in (\ref{CS2}) are redundant given (\ref{TRB4}) and (\ref{TRB1}). Only cut-set bounds in (\ref{CS3}) remain useful. Then, we obtain the following theorem.
\begin{theorem}
\label{OB}
The capacity region $\mathcal{C}$ of the DYC is outer bounded by $\overline{\mathcal{C}}$, where
\begin{align}
\label{OBE}
\overline{\mathcal{C}}\triangleq\left\{
\begin{array}{rl}
\vec{R}\in\mathbb{R}_+^6|& \text{$\vec{R}$ satisfies \eqref{TRB1}-\eqref{TRB6} and \eqref{CS3}} 
\end{array}
\right\}.
\end{align}
\end{theorem}

In the next section, we show that this outer bound is achievable and characterize the capacity region of the DYC.

\section{Achievability}
\label{Achievability}

\begin{figure}
\centering
\subfigure[Relay levels in the uplink.]{
\includegraphics[height=0.35\columnwidth]{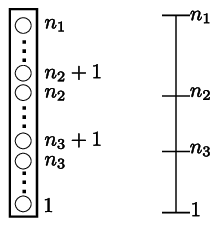}
\label{RelayLevelsU}
}
\hspace{0.5cm}
\subfigure[Relay levels in the downlink.]{
\includegraphics[height=0.35\columnwidth]{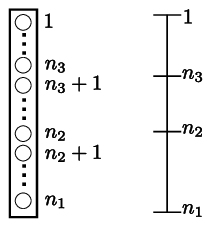}
\label{RelayLevelsD}
}
\caption{Enumeration of levels at the relay. Levels 1 to $n_3$ are seen by all three users, levels $n_3+1$ to $n_2$ are seen by users 1 and 2, while the remaining levels are seen only by user 1.}
\end{figure}

In what follows, we enumerate the levels at the relay as shown in Fig. \ref{RelayLevelsU} and \ref{RelayLevelsD}. In the uplink, the lowest level is level 1 and the highest is level $q=n_1$ while in the downlink, the lowest level is $q=n_1$ and the highest is level 1. In both the uplink and downlink, levels $\{1,\dots,n_3\}$ are accessible by all 3 users, levels $\{n_3+1,\dots, n_2\}$ are accessible by users 1 and 2, and levels $\{n_2+1,\dots,n_1\}$ are accessible by user 1 only. This makes this enumeration convenient for describing the achievable scheme. We represent the levels at the relay by a line segment as shown in Fig. \ref{RelayLevelsU} and \ref{RelayLevelsD}.

We start by showing that any rate tuple $\vec{R}\in\mathbb{N}^6\cap\overline{\mathcal{C}}$ is achievable. $\vec{R}\in\overline{\mathcal{C}}$ implies that it satisfies the bounds \eqref{TRB1}-\eqref{TRB6} and \eqref{CS3}. Now, we have to show that we can use the signal levels at the relay wisely to achieve $\vec{R}$. Our scheme uses three different strategies to cover three different cases of communication. These cases are as follows:
\begin{itemize}
\item[A)] \textbf{Bi-directional:} There exist users that want to establish bi-directional communication. That is, $R_{jk}$ and $R_{kj}$ are both non-zero for some $j,k\in\{1,2,3\}$, $j\neq k$. 
\item[B)] \textbf{Cyclic:} Users only want to establish cyclic communication. That is, $R_{jk}$, $R_{kl}$, and $R_{lj}$ are non-zero while $R_{kj}=R_{lk}=R_{jl}=0$ for some distinct $j,k,l\in\{1,2,3\}$.
\item[C)]\textbf{Uni-directional:} Neither case A) nor B) hold.
\end{itemize}

For a given $\vec{R}\in\mathbb{N}^6\cap\overline{\mathcal{C}}$, the suggested scheme starts with a bi-directional communication strategy if case A) holds, which operates at a rate of two bits per relay level as we see next. After this step, some rates are already achieved and the residual rate vector is called $\vec{R}'$. We also have a reduced DYC obtained by removing the already occupied levels. It remains to achieve $\vec{R}'$ which has at least three zero components over this reduced DYC. Now we use the cyclic strategy if case B) holds, which operates at a rate of 3/2 bits per level. After this step, the residual rate vector, denoted $\vec{R}''$ belongs to case C), and we use the uni-directional strategy to achieve it, which operates at a rate of 1 bit per level. These strategies are depicted in Fig. \ref{RL} and explained in the next subsections.

\subsection{Bi-directional communication over the DYC}
\label{BDC}
In the bi-directional communication strategy, 2 bits, one from each user involved in bi-directional communication, consume one level at the relay. Let\footnote{$a$, $b$, or $c$ can have zero value. If all are zero, then this strategy does not exist and we start with case B) instead.}
\begin{align*}
a=\min\{R_{12},R_{21}\},\ b=\min\{R_{13},R_{31}\},\ c=\min\{R_{23},R_{32}\}.
\end{align*}
Users 1 and 2 use levels $\{n_2-a+1,\dots,n_2\}$ in a manner similar to the deterministic bi-directional relay channel \cite{AvestimehrSezginTse}. That is, users 1 and 2 send binary vectors, say $\vec{x}_{12}$ and $\vec{x}_{21}$ from $\mathbb{F}_2^a$, on relay levels $\{n_2-a+1,\dots,n_2\}$. The relay obtains $\vec{x}_{12}\oplus\vec{x}_{21}$ and sends it back on the same levels\footnote{with ``same levels", we mean levels with the same indexes.}. Users 1 and 2 then, knowing their transmit vector, calculate their desired information from the $\vec{x}_{12}\oplus\vec{x}_{21}$. Similarly, users 1 and 3 use levels $\{1,\dots,b\}$ and users 2 and 3 use levels $\{b+1,\dots,b+c\}$.

This strategy works if we have enough levels at the relay for all $a+b+c$ bi-directional streams, i.e., $a+b+c\leq n_2$ and $b+c\leq n_3$ (cf. Fig. \ref{RelayLevels1}). But since $\vec{R}\in\overline{\mathcal{C}}$ then
\vspace{-0.2cm}
\begin{align}
b+c&\leq R_{13}+R_{23}\stackrel{\eqref{CS3}}{\leq} n_3\\
a+b+c&\leq R_{12}+R_{13}+R_{23}\stackrel{\eqref{TRB5}}{\leq} n_2,
\end{align}
and thus, the levels at the relay are sufficient for this strategy to work. Now, we need to achieve 
\begin{align}
\vec{R}'&=(R_{12}-a,R_{13}-b,R_{21}-a,R_{23}-c,R_{31}-b,R_{32}-c)\nonumber\\
\label{Rp}
&\triangleq(R_{12}',R_{13}',R_{21}',R_{23}',R_{31}',R_{32}'),
\end{align}
where either $R_{jk}'=0$ or $R_{kj}'=0$, over DYC$(n_1',n_2',n_3')$ with \begin{align}
n_1'&=n_1-a-b-c,\\
\label{n2p}
n_2'&=n_2-a-b-c.
\end{align}
Recall that the bi-directional communication between users 1 and 2 use relay levels $\{n_2-a+1,\dots,n_2\}$. This does not consume levels in $\{1,\dots,n_3\}$ if $n_2-n_3\geq a$, in which case $n_3'=n_3-b-c$. Otherwise, then $a-n_2+n_3$ levels in $\{1,\dots,n_3\}$ are used for this communication and in this case $n_3'=n_3-b-c-(a-n_2+n_3)=n_2'$. Therefore,
\begin{align}
n_3'=\min\{n_3-b-c,n_2'\}.
\end{align}
The remaining non-zero components of $\vec{R}'$ can represent cyclic communication as in case B), or uni-directional communication as in case C).

\subsection{Cyclic communication over the DYC}
\label{CC}
If case B) holds, then users want to communicate in a cyclic manner over a DYC$(n_1',n_2',n_3')$. There are two possible cycles, either $1\to2\to3\to1$ or $1\to3\to2\to1$. Let
\begin{align}
\label{D}
d&=\min\{R_{12}',R_{23}',R_{31}'\},\\
\label{E}
e&=\min\{R_{13}',R_{32}',R_{21}'\}.
\end{align}
Notice that either $d$ or $e$ must be zero, since otherwise, we would have bi-directional communication which would have been taken care of in the previous step in Section \ref{BDC}. If both are zero, then this strategy is skipped and case C) is considered instead. It is easy to see that
\begin{align}
\label{R1}
d>0&\Rightarrow e=0,\ a=R_{21},\ b=R_{13},\ c=R_{32},\\
\label{R2}
e>0&\Rightarrow d=0,\ a=R_{12},\ b=R_{31},\ c=R_{23}.
\end{align}

\begin{figure*}[t]
\centering
\subfigure[Bi-directional~communication.]{
\includegraphics[height=0.4\columnwidth]{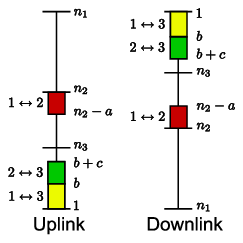}
\label{RelayLevels1}
}
\hspace{0.5cm}
\subfigure[Cyclic communication: $1\to2\to3\to1$.]{
\includegraphics[height=0.4\columnwidth]{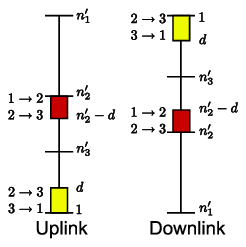}
\label{RelayLevels2}
}
\hspace{0.5cm}
\subfigure[Cyclic communication: $1\to3\to2\to1$.]{
\includegraphics[height=0.4\columnwidth]{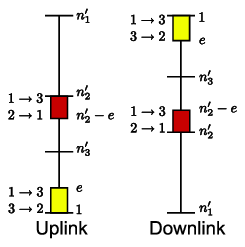}
\label{RelayLevels3}
}
\hspace{0.5cm}
\subfigure[Uni-directional communication.]{
\includegraphics[height=0.4\columnwidth]{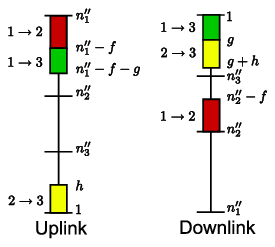}
\label{RelayLevels4}
}
\caption{Assignment of levels at the relay for different communication strategies.}
\label{RL}
\end{figure*}

Let us consider the first case (\ref{R1}). Thus $R_{12}'$, $R_{23}'$ and $R_{31}'$ are non-zero and the other rates are zero, i.e. $e=0$. Let the transmit vectors of users 1, 2, and 3 be $\vec{x}_{12}$, $\vec{x}_{23}$ and $\vec{x}_{31}$ all in $\mathbb{F}_2^d$. Users 1 and 2 send $\vec{x}_{12}$ and $\vec{x}_{23}$ on relay levels $\{n_2'-d+1,\dots,n_2'\}$. Users 2 also repeats $\vec{x}_{23}$ on relay levels $\{1,\dots,d\}$ together with user 3 which sends $\vec{x}_{31}$ on the same levels. The relay receives $\vec{x}_{12}\oplus\vec{x}_{23}$ and $\vec{x}_{23}\oplus\vec{x}_{31}$ and sends them back on the same levels. Users 1 and 2 receive $\vec{w}_1=\vec{x}_{12}\oplus\vec{x}_{23}$ and $\vec{w}_2=\vec{x}_{23}\oplus\vec{x}_{31}$ since all bits are sent on levels below $n_2'$. Here, we need $2d\leq n_2'$ (cf. Fig. \ref{RelayLevels2}). Then, knowing $\vec{x}_{12}$, user 1 extracts $\vec{x}_{23}$ from $\vec{w}_{1}$ and afterwards extracts $\vec{x}_{31}$ from $\vec{w}_{2}$. User 2, knowing $\vec{x}_{23}$, extracts $\vec{x}_{12}$ from $\vec{w}_{1}$. If $d\leq n_3'$ then user 3 receives $\vec{w}_{2}$ and, knowing $\vec{x}_{31}$, extracts $\vec{x}_{23}$ from $\vec{w}_2$. 

As long as $d\leq n_3'$ and $2d\leq n_2'$, then $2d$ levels are sufficient for communicating $3d$ bits, for an average of 3/2 bits per level. But these inequalities hold as long as $\vec{R}\in\overline{\mathcal{C}}$ since
\begin{align}
d&\stackrel{\footnotesize{(\ref{D})}}{\leq} R_{31}'\stackrel{\footnotesize{(\ref{Rp})}}{=}R_{31}-b\stackrel{\footnotesize{(\ref{CS3})}}{\leq} n_3-R_{32}-b\stackrel{\footnotesize{(\ref{R1})}}{=}n_3-b-c\nonumber\\
2d&\stackrel{\footnotesize{(\ref{D})}}{\leq} R_{31}'+R_{23}'\stackrel{\footnotesize{(\ref{Rp})}}{=}R_{31}+R_{23}-b-c\\
&\stackrel{\footnotesize{(\ref{TRB4})}}{\leq} n_2-R_{21}-b-c\stackrel{\footnotesize{(\ref{R1})}}{=}n_2-a-b-c\stackrel{\footnotesize{(\ref{n2p})}}{=}n_2'.
\end{align}
Thus $d\leq\min\{n_3-b-c,n_2'\}=n_3'$ and $2d\leq n_2'$, and therefore, there are enough levels in the DYC$(n_1',n_2',n_3')$ for serving all bits of cyclic communication. For the second possibility, i.e. $1\to3\to2\to1$, a similar strategy can be used. User 1 repeats a bit on two levels. One level must be in $\{1,\dots,n_2'\}$ and the other in $\{1,\dots,n_3'\}$. Using (\ref{CS3}), (\ref{TRB5}) and (\ref{TRB3}), we can show that the levels at the relay $(n_1',n_2',n_3')$ are sufficient for this communication (Keeping in mind that in this case $a=R_{12}$, $b=R_{31}$, $c=R_{23}$, and $d=0$). The assignment of the levels at the relay in this case is shown in Fig. \ref{RelayLevels3}. After this stage, the rate tuple that still needs to be achieved is
\begin{align}
\vec{R}''&=(R_{12}'-d,R_{13}'-e,R_{21}'-e,R_{23}'-d,R_{31}'-d,R_{32}'-e)\nonumber\\
\label{Rpp}
&\triangleq(R_{12}'',R_{13}'',R_{21}'',R_{23}'',R_{31}'',R_{32}''),
\end{align}
over a DYC$(n_1'',n_2'',n_3'')$ where 
\begin{align}
n_1''&=n_1'-2d-2e,\\
n_2''&=n_2'-2d-2e.
\end{align}
If $d+e\leq n_2'-n_3'$ then $n_3''=n_3'-d-e$, otherwise, $n_3''=n_3'-d-e-(d+e-n_2'+n_3')=n_2''$. Thus 
\begin{align}
n_3''=\min\{n_3'-d-e,n_2''\}.
\end{align}
% After considering cases A) and B), only uni-directional communication remains.

\subsection{Uni-directional communication}
\label{AUC}
Finally, we need to achieve $\vec{R}''$ with at least 3 zero components over a DYC$(n_1'',n_2'',n_3'')$. The non-zero components of $\vec{R}''$ do not represent bi-directional nor cyclic communication. We have 6 different cases, one of which is described in details, and the rest follow similarly. Here, each bit consumes one relay level.

We consider the following case: $R_{21}'',R_{31}'',R_{32}''=(0,0,0)$. Let $R_{12}''=f$, $R_{13}''=g$, $R_{23}''=h$ where $f,g,h\geq0$ and let $\vec{x}_{12}\in\mathbb{F}_2^f$, $\vec{x}_{13}\in\mathbb{F}_2^g$, and $\vec{x}_{23}\in\mathbb{F}_2^h$ denote the binary vectors to be communicated. In the uplink, user 1 uses levels $\{n_1''-f+1,\dots,n_1''\}$ to send $\vec{x}_{12}$ and levels $\{n_1''-f-g+1,\dots,n_1''-f\}$ to send $\vec{x}_{13}$, and user 2 uses levels $\{1,\dots,h\}$ to send $\vec{x}_{23}$ to the relay. The relay then forwards $\vec{x}_{13}$ on levels $\{1,\dots,g\}$, $\vec{x}_{23}$ on levels $\{g+1,\dots,g+h\}$, and $\vec{x}_{12}$ on levels $\{n_2''-f+1,\dots,n_2''\}$ (cf. Fig. \ref{RelayLevels4}). This works for communicating all $f+g+h$ bits in the uplink if: $R_{23}''\leq n_2''$ and $R_{12}''+R_{13}''+R_{23}''\leq n_1''$, and in the downlink if: $R_{23}''\leq n_3''$, $R_{23}''+R_{13}''\leq n_3''$, and $R_{12}''+R_{13}''+R_{23}''\leq n_2''$. Combining, we get 
\begin{align}
\label{InEq1}
R_{23}''+R_{13}''&\leq n_3''\\
\label{InEq2}
R_{12}''+R_{13}''+R_{23}''&\leq n_2''.
\end{align}
These inequalities are satisfied as long as $\vec{R}\in\overline{\mathcal{C}}$. Consider the first inequality (\ref{InEq1}),
\begin{align}
R_{23}''+R_{13}''&\stackrel{\footnotesize{(\ref{Rpp})}}{=}R_{23}'+R_{13}'-d-e\\
\label{R23ppPR13pp2}
&\stackrel{\footnotesize{(\ref{Rp})}}{=}R_{23}+R_{13}-b-c-d-e\\
&\stackrel{\footnotesize{(\ref{CS3})}}{\leq} n_3-b-c-d-e.
\end{align}
Recall that either $d$ or $e$ must be zero. If $e=0$ then $R_{23}''+R_{13}''\stackrel{\footnotesize{(\ref{R23ppPR13pp2})}}{=}R_{23}+R_{13}-b-c-d\stackrel{\footnotesize{(\ref{TRB5})}}{\leq} n_2-R_{12}-b-c-d\stackrel{\footnotesize{(\ref{D})}}{\leq} n_2'-2d$. If $e>0$, then $d=0$, $a=R_{12}$, and $c=R_{23}$ by (\ref{R2}) and thus
$R_{23}''+R_{13}''\stackrel{\footnotesize{(\ref{R23ppPR13pp2}}}{=}R_{23}+R_{13}-b-c-e\stackrel{\footnotesize{(\ref{R2})}}{=} R_{13}-b-e\stackrel{\footnotesize{(\ref{TRB1})}}{\leq} n_2-R_{12}-R_{32}-b-e\stackrel{\footnotesize{(\ref{E})}}{\leq} n_2'-2e$, where the last inequality follows from (\ref{E}) since $e\leq R_{32}'=R_{32}-c$ by (\ref{Rp}). Thus, since either $d=0$ or $e=0$ we get
\begin{align*}
R_{23}''+R_{13}''\leq \min\{n_3-b-c-d-e, n_2'-2d-2e\}=n_3''.
\end{align*}
Consider now the second inequality (\ref{InEq2}). From (\ref{Rp}) and (\ref{Rpp}) we have $R_{12}''+R_{13}''+R_{23}''\leq R_{12}+R_{13}+R_{23}-a-b-c-2d-e$. If $e=0$ then $R_{12}''+R_{13}''+R_{23}''\leq R_{12}+R_{13}+R_{23}-a-b-c-2d\stackrel{\footnotesize{(\ref{TRB5})}}{\leq} n_2-a-b-c-2d$. Otherwise, if $e>0$ then $d=0$ and using $c=R_{23}$ from (\ref{R2}) we get $R_{12}''+R_{13}''+R_{23}''\leq R_{12}+R_{13}-a-b-e\stackrel{\footnotesize{(\ref{TRB1})}}{\leq} n_2-R_{32}-a-b-e\stackrel{\footnotesize{(\ref{E})}}{\leq} n_2-a-b-c-2e$, where the last inequality follows from (\ref{E}) since $e\leq R_{32}'=R_{32}-c$ by (\ref{Rp}). We obtain
\begin{align}
R_{12}''+R_{13}''+R_{23}''&\leq n_2'-2d-2e=n_2''.
\end{align}
As a result, both inequalities \eqref{InEq1} and \eqref{InEq2} are satisfied, and there exist enough levels for communicating all $f+g+h$ bits. The remaining cases of uni-directional communication are similar to the case studied above. Consequently, after assigning levels for bi-directional communication and for cyclic communication, enough levels remain to communicate all the remaining bits in $\vec{R}''$. We obtain the following theorem.

\begin{theorem}
\label{IB}
Every rate tuple $\vec{R}\in\mathbb{N}^6\cap\overline{\mathcal{C}}$ is achievable.
\end{theorem}
\begin{proof}
Using the schemes described in Sections \ref{BDC}, \ref{CC}, and \ref{AUC} we can achieve any integer vector $\vec{R}$ that belongs to $\overline{\mathcal{C}}$ and the result follows.
\end{proof}

It was shown in \cite{AvestimehrKhajehnejadSezginHassibi} that studying a multi-pair bi-directional relay network considered over $Q$ symbol extensions ($Q$ time slots) is the same as the original network with channel gains multiplied by $Q$. Same statement holds here. We can think of a DYC$(n_1,n_2,n_3)$ over $Q$ time slots as a DYC$(Qn_1,Qn_2,Qn_3)$. We obtain the following theorem.

\begin{theorem}
The capacity region $\mathcal{C}$ of the DYC is $\overline{\mathcal{C}}$.
\end{theorem}
\begin{proof}
To achieve $\overline{\mathcal{C}}$, it is sufficient to achieve its corner points. All other points in $\overline{\mathcal{C}}$ can be then achieved by time sharing between different corner points. Let us show that all corner points are achievable. Since all inequalities representing the boundary of the outer bound $\overline{\mathcal{C}}$, i.e. (\ref{TRB1})-(\ref{TRB6}) and (\ref{CS3}), have integer coefficients, then all its corner points are fractional. Consider a corner point $$\vec{R}=\left(\frac{P_{12}}{Q_{12}},\frac{P_{13}}{Q_{13}},\frac{P_{21}}{Q_{21}},\frac{P_{23}}{Q_{23}},\frac{P_{31}}{Q_{31}},\frac{P_{32}}{Q_{32}}\right),$$
where $P_{jk},Q_{jk}\in\mathbb{N}$ for all $j,k\in\{1,2,3\}$, $j\neq k$. This corner point is achievable as follows. Use $Q$ time slots to achieve the rate tuple $Q\vec{R}$ where $Q=\prod_{j=1}^3\prod_{k=1,\ k\neq j}^3Q_{jk},$
over a DYC$(Qn_1,Qn_2,Qn_3)$. Since $\vec{R}\in\overline{\mathcal{C}}$ then $Q\vec{R}\in\overline{\mathcal{C}'}$ where $\overline{\mathcal{C}'}$ is the outer bound given in Theorem \ref{OB} for a DYC$(Qn_1,Qn_2,Qn_3)$. Moreover, $Q\vec{R}\in\mathbb{N}^6$. Thus $Q\vec{R}\in\mathbb{N}^6\cap\overline{\mathcal{C}'}$ which means that it is achievable according to Theorem \ref{IB}. But $Q\vec{R}$ is achievable in a DYC$(Qn_1,Qn_2,Qn_3)$ implies that $\vec{R}$ is achievable in a DYC$(n_1,n_2,n_3)$. Therefore all corner points of $\overline{\mathcal{C}}$ are achievable, and the statement of the theorem follows.
\end{proof}

Notice that the capacity region given by $\overline{\mathcal{C}}$ is not symmetric in general. This is due to the asymmetry in the channel owing to the different values of $n_1$, $n_2$ and $n_3$.

\section{Summary}
\label{Summary}
We studied the linear shift deterministic Y-channel and obtained its capacity region. We first obtained an outer bound by considering genie aided upper bounds on achievable rates in addition to the cut-set bounds. Then we showed that this outer bound is achievable. The achievability scheme assigns levels at the relay in a way that suffices for achieving any integer rate tuple in the outer bound. Namely, three schemes are used: bi-directional, cyclic, and uni-directional communication. The levels at the relay are shown to be enough to perform this assignment as long as the rate tuple is within the outer bound. Then, the corner points of the outer bound are shown to be achievable by using symbol extensions. Thus the outer bound is achievable and is the capacity region of the linear shift deterministic Y-channel.

\bibliography{myBib}

\end{document}